\documentclass[sigconf, natbib=true]{acmart}
\usepackage[dvipsnames, svgnames, x11names,table,xcdraw]{xcolor}
\usepackage{multirow, soul}
\usepackage{enumitem}
\usepackage{tabularx}
\usepackage{algorithm}
\usepackage{algpseudocode}

\AtBeginDocument{%
  }

\setcopyright{acmlicensed}
\copyrightyear{2018}
\acmYear{2018}
\acmDOI{XXXXXXX.XXXXXXX}
\acmConference[Conference acronym 'XX]{Make sure to enter the correct
  conference title from your rights confirmation email}{June 03--05,
  2018}{Woodstock, NY}
\acmISBN{978-1-4503-XXXX-X/2018/06}

\begin{document}

\title{Breaking User-Centric Agency: A Tri-Party Framework for Agent-Based Recommendation}



\author{Yaxin Gong}
\email{gyx2022@mail.ustc.edu.cn}
\affiliation{%
 \institution{University of Science and Technology of China}
 \city{Hefei}
 \country{China}}

\author{Chongming Gao}
\email{chongminggao@ustc.edu.cn}
\affiliation{%
 \institution{University of Science and Technology of China}
 \city{Hefei}
 \country{China}}

\author{Chenxiao Fan}
\email{simonfan@mail.ustc.edu.cn}
\affiliation{%
 \institution{University of Science and Technology of China}
 \city{Hefei}
 \country{China}}

\author{Haoyan Liu}
\email{liuhaoyan@ustc.edu.cn}
\affiliation{%
 \institution{University of Science and Technology of China}
 \city{Hefei}
 \country{China}}

\author{Wenjie Wang}
\email{wenjiewang96@gmail.com}
\affiliation{%
 \institution{University of Science and Technology of China}
 \city{Hefei}
 \country{China}}

\author{Jianshan Sun}
\email{sunjs9413@hfut.edu.cn}
\affiliation{%
 \institution{Hefei University of Technology}
 \city{Hefei}
 \country{China}}

\author{Yangyang Li}
\email{liyangyang@live.com}
\affiliation{%
 \institution{Academy of Cyber}
 \city{Beijing}
 \country{China}}

\author{Fuli Feng}
\email{fulifeng93@gmail.com}
\affiliation{%
 \institution{University of Science and Technology of China}
 \city{Hefei}
 \country{China}}

\author{Xiangnan He}
\email{xiangnanhe@gmail.com}
\affiliation{%
 \institution{University of Science and Technology of China}
 \city{Hefei}
 \country{China}}

\renewcommand{\shortauthors}{Gong et al.}


\begin{abstract}
Large language models (LLMs) have spurred interest in agent-based recommender systems, yet most agentic approaches remain user-centric: items stay passive entities whose exposure is a by-product of relevance ranking, which exacerbates exposure concentration and long-tail under-representation.
We break this user-centric allocation of agency with a \textbf{Tri}-party LLM-agent \textbf{Rec}ommendation framework (TriRec). Responsibility is split deliberately: Stage~1 has each item generate self-promotion conditioned on the target user, which lowers cold-start barriers, while the exposure budget stays with the platform, whose Stage~2 sequential re-ranker balances relevance, item utility, and exposure fairness.
On four public datasets TriRec improves accuracy, fairness, and item-level utility, with the accuracy gain significant on three of the four. A three-arm ablation at $50$ candidates separates two levels of the mechanism on items that received no exposure during training: \emph{self-promotion} drives the accuracy gain, and \emph{conditioning} it on the target user adds further exposure, together raising these items' share of top-ranked exposure by $43.6\%$ relative. Restricting promotions to catalogue-verifiable attributes cuts strong exaggeration to $0.5\%$/$2.0\%$ on two datasets while retaining $88.6\%$/$92.0\%$ of the accuracy gain. Our code is available at \url{https://github.com/Marfekey/TriRec}.
\end{abstract}




\maketitle

\begin{figure}[h]
  \centering
  \includegraphics[width=\linewidth]{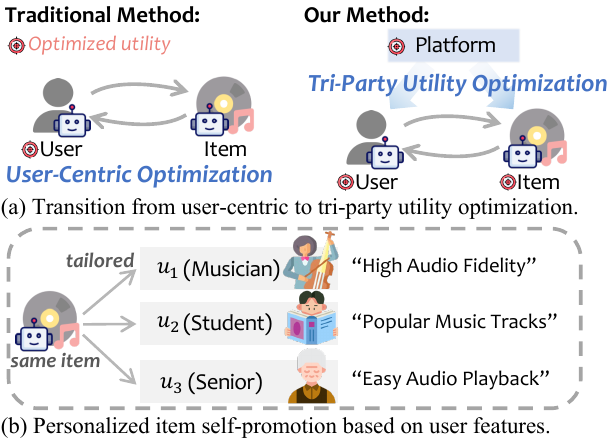}
  \caption{Illustration of tri-party agentic recommendation and personalized item self-promotion.}
  \label{fig:intro}
\end{figure}
\section{Introduction}


Large language models (LLMs) have demonstrated remarkable capabilities across a wide range of domains~\cite{tunable,oasis,lyu-etal-2024-llm,10.1145/3757925}, fundamentally reshaping how intelligent decision-making systems reason~\cite{zhu-etal-2025-llm-based,2compe} and interact with users~\cite{usersim, user_sim_tois}. In recommender systems, these advances have stimulated growing interest in LLM-based agent paradigms, where agents model user behaviors through language-driven interaction and reasoning~\cite{afl,rec4agentverse}, enabling more expressive and interpretable preference representations compared to traditional collaborative filtering approaches~\cite{agent4rec,macrec,ecommerce}.

However, despite these promising developments, most existing agent-based recommender methods largely mirror human-centered decision-making patterns~\cite{interecagent,doIT} and adopt a predominantly user-centric modeling paradigm~\cite{dyta4rec,recmind,hybrid_macrs,iagent}. They primarily optimize user utility while treating items as passive functional entities~\cite{agentsociety,reasonrec}, thereby failing to adequately account for the interests of other critical stakeholders, such as content providers and platform operators~\cite{simuser,agentcf,starec,user_sim_tois}. Such a user-centric decision paradigm is inherently unsustainable in multi-sided marketplaces. In particular, the pervasive Matthew effect concentrates exposure on already popular items~\cite{sprec}, leaving long-tail creators with limited visibility~\cite{makov1, makov2}. This imbalance weakens creator incentives, leading to creator churn and a subsequent decline in high-quality content production~\cite{dualrec}. Moreover, excessive optimization for short-term user engagement exacerbates a winner-take-all dynamic, gradually homogenizing the content pool. Ultimately, this erosion of supply-side diversity undermines user experience and harms the platform's long-term ecosystem health.

Although some recent methods introduce item agents or content agents~\cite{agentcf,creatorag,rec4agentverse, agentcf++}, these agents typically function only as information carriers or attribute descriptors. Their role remains largely auxiliary, serving to better satisfy user preferences rather than actively pursuing reasonable exposure or long-term benefits on behalf of the items themselves.

To address this limitation at its root, we propose a \textbf{Tri}-party LLM-agent \textbf{Rec}ommendation framework (TriRec) that explicitly aligns the interests of users, items, and the platform, as illustrated in Figure~\ref{fig:intro}. The framework adopts a disentangled two-stage architecture that preserves user experience while letting items describe themselves through personalized self-promotion, i.e., generating user-conditioned content that highlights how each item matches the target user. At the platform level, global ranking is further regulated by incorporating system-level constraints that govern exposure allocation within each ranking round.

Stage 1: Generative item self-promotion. Item agents generate self-promotion instead of being treated as fixed candidates: each item produces a description tailored to the target user's interest preference, while exposure remains allocated by the platform rather than won by the item. For example, the same CD player might emphasize ``high audio fidelity'' to musicians and ``easy audio playback'' to seniors. The platform's Stage~2 re-ranker then turns a better match into exposure for long-tail items.

Stage 2: Platform-led multi-objective re-ranking. 
After Stage 1 generates a high-quality candidate list, the platform performs sequential re-ranking to balance multiple objectives. Specifically, for each candidate item, the platform re-ranker considers immediate user relevance, platform-level fairness and expected item utility.
The platform makes ranking decisions one position at a time; this sequential strategy lets each ranking choice account for exposure already accumulated earlier in the evaluation stream.

Notably, incorporating item self-promotion increased both the exposure share and the expected click-through of items that carry no interaction history, and it improved recommendation accuracy on the user side; conditioning the promotion on the target user contributed to the former rather than the latter. These results demonstrate that our framework can serve the objectives of users, items, and the platform, while making the division of labour among its components explicit.

Our main contributions are as follows:
\begin{itemize}[leftmargin=*]
    \item We identify fundamental limitations of user-centric modeling and introduce a tri-party LLM-agent framework that coordinates users, items, and the platform through agent-based exposure regulation.
    \item We propose a two-stage pipeline: (i) item-side self-promotion, in which items generate user-conditioned content to improve matching and alleviate cold start, and (ii) platform-led multi-objective re-ranking with exposure modulation to balance tri-party interests.
    \item Experiments on four datasets show TriRec outperforms existing baselines in accuracy, fairness, and item utility, with the accuracy gain statistically significant on three of the four datasets. A controlled ablation on a five-fold larger pool, with significance testing over items that received no exposure during training, separates two levels of the mechanism: \emph{self-promotion} drives most of the gain, and \emph{conditioning} it on the target user adds further exposure.
\end{itemize}

\section{Related Work}
\subsection{LLM-based Agentic Recommendation}
The reasoning capabilities of LLMs have reshaped intelligent decision-making, stimulating interest in LLM-based agent paradigms for recommendation~\cite{Gao_survey_2026, user_sim_tois}. Unlike traditional filtering, these agents model complex behaviors through language-driven reasoning, enabling more expressive preference representations~\cite{agentcf,NextQuill}. One prominent direction employs agents as high-fidelity surrogates of human behavior~\cite{simuser, oasis}: Agent4Rec~\cite{agent4rec} and DyTA4Rec~\cite{dyta4rec} incorporate sociological traits and dynamic profile updates to capture interest evolution, while PUMA~\cite{dontstart} migrates personalized prompt assets across evolving model architectures. Others use agents as intelligent interfaces that improve decision quality through deliberate reasoning~\cite{recmind, interecagent, iagent}; STARec~\cite{starec} and ReasonRec~\cite{reasonrec} add reasoning mechanisms to mitigate hallucinations, and LatentR3~\cite{R3} replaces explicit text generation with reinforced latent tokens for efficiency.

Decentralized multi-agent architectures instead decompose complex tasks into specialized roles or model bidirectional dynamics~\cite{macrec, hybrid_macrs,creatorag}. AgentCF~\cite{agentcf} and Rec4Agentverse~\cite{rec4agentverse} introduce item agents that interact autonomously with user agents for preference refinement, and the bi-learning planner~\cite{bi-learning} coordinates macro-level guidance for sustained engagement. Two mechanisms are nevertheless absent: the item-side text these agents produce is not conditioned on the specific user being served, and no platform-side exposure control is paired with it, so exposure remains a by-product of relevance ranking. TriRec supplies both, and \S\ref{sec:coldstart} measures what each contributes.

\subsection{Creator-side Recommendation}
User-centric recommendation concentrates most exposure on a few popular items, leaving many creators invisible; this hinders the creator experience and ultimately depresses content production and platform engagement~\cite{ipsrec, dualrec}. Early methodologies synchronized user and provider representations via co-learning~\cite{ipsrec}, or provided counterfactual interpretability for the provider side, as in PRINCE~\cite{prince}.

Recent work ``mirrors'' user-centric techniques: DualRec~\cite{dualrec} treats items as active queries to identify suitable users and tackle the user availability challenge in dual-target optimization, while generative approaches such as HLLM-Creator~\cite{hllm_creator} and FlyThinker~\cite{ThinkWhileGenerating} leverage LLMs to augment item content and dynamically guide long-form personalized responses. However, these frameworks still view creators as passive targets for algorithmic optimization or static subjects of content generation, lacking explicit modeling of autonomous agency. Different from these passive approaches, our work empowers creators with agentic capabilities, enabling tailored promotion of their content to users.

\subsection{Platform Fair Ranking}
The pervasive Matthew effect leads to exposure concentration, leaving vast long-tail products invisible and causing cold-start problems~\cite{fairness_exposure, pmmf}, so platforms must intervene to reconcile accuracy with multi-stakeholder fairness~\cite{tfrom}. Classical research formulates exposure allocation as constrained optimization~\cite{fairness_exposure, cpfairrank} or learns debiased representations~\cite{profairrec, tfrom} to neutralize provider-side biases; SPRec~\cite{sprec} adds self-play to mitigate filter bubbles without extra data.

Another dominant paradigm applies post-hoc re-ranking under social welfare criteria. LTP-MMF~\cite{ltp_mmf} mitigates the deleterious feedback loop effect, the bi-learning planner~\cite{bi-learning} manages macro-level guidance for long-term user engagement, and multi-agent social choice frameworks such as SCRUF-D~\cite{scruf_d} arbitrate conflicting fairness criteria. While these methods move toward decentralized decision-making, they struggle to perceive fine-grained item-level signals. In this work, we formulate platform regulation as re-ranking that adaptively coordinates relevance, fairness, and item utility.


\section{Problem Formulation}

We formally model the recommendation problem addressed by TriRec from a multi-stakeholder and agentic perspective. Unlike conventional settings that optimize user relevance alone, our framework explicitly accounts for the potentially conflicting objectives of users, items, and the platform, and introduces the agent-based preference interaction mechanism underlying Stage~1 candidate construction.

\subsection{Multi-Stakeholder Utility Modeling}
\label{sec:problem_formulation}
We therefore define utility functions for users, items, and the platform, which together characterize the overall optimization target of the recommendation process.

\begin{itemize}[leftmargin=*]

    \item \textbf{User Utility.} 
    Users primarily seek high relevance and engagement quality~\cite{NextQuill}. 
    At the platform stage, we define the user-side utility on a per-item basis:
    \begin{equation}
    U_{\text{user}}(u, i) = r_{\text{LLM}}(u,i),
    \end{equation}
    where $r_{\text{LLM}}(u,i)$ is the relevance score predicted by the user agent.
    
    \item \textbf{Item Utility.} 
    Items aim to obtain effective exposure and long-term visibility across recommendation opportunities.
    We quantify the realized item-side gain $U_{\text{item}}$ using the \textit{Expected Item Utility (EIU)}:
    \begin{equation}
        EIU(u,i) = v(\text{rank}_u(i)) \cdot \text{CTR}(u,i),
    \label{uitem}
    \end{equation}
    where $v(\cdot)$ models position-dependent exposure probability following a logarithmic decay, and $\text{CTR}(u,i)$ denotes 
    the predicted click-through probability.
    In our implementation, $\text{CTR}(u,i)$ is estimated by a sigmoid 
    transformation of the cosine similarity between user and item 
    semantic embeddings, and cumulative item utility is aggregated 
    across users and recommendation rounds. We validate this proxy 
    against held-out purchases in \S\ref{sec:itemutility}.

    \item \textbf{Platform Utility.} 
    The platform aims to regulate exposure allocation and mitigate group-level bias across the recommended items.
    We denote the platform utility as $U_{\text{platform}}(\Pi)$, where $\Pi$ aggregates ranking outcomes over users and time.
    In practice, platform-level fairness is quantified using Distributional Group Unfairness (DGU) and Maximal Group Unfairness (MGU)~\cite{flower}, which measure the discrepancy between the exposure distribution of item groups in top-$K$ recommendation results and their historical distribution in the training data. Concretely, all items are sorted by their training-set exposure count and partitioned into eight equal-sized groups; DGU is the total-variation distance $\tfrac{1}{2}\sum_g|p_{\text{rec}}(g)-p_{\text{train}}(g)|$ between the recommended and training group distributions, and MGU is the maximum per-group gap $\max_g|p_{\text{rec}}(g)-p_{\text{train}}(g)|$.

\end{itemize}

Based on the above utility definitions, TriRec realizes recommendation through a two-stage agentic pipeline. Stage~1 constructs relevance-oriented candidate rankings by modeling fine-grained user--item semantic alignment via agent-based interaction and personalized item self-promotion, producing a preference backbone without exposure regulation. Stage~2 then treats recommendation as a sequential control problem: the platform re-ranker observes the system state, including historical exposure and relevance rankings, and re-ranks to jointly optimize user utility, item exposure utility, and fairness over time. This decoupled design separates semantic preference modeling from long-term exposure regulation.


\subsection{Agent-based Preference Interaction}
\label{sec:agent_interaction}
Following AgentCF~\cite{agentcf}, we adopt an agent-based recommendation paradigm to model fine-grained user--item preference alignment. Through offline replay of historical interaction episodes, user agents accumulate preference memory while item agents build item-side semantic memory capturing their appeal across user groups; the item-side memory is further updated online during Stage~1 (Eq.~\ref{equ:mem_update}).

Formally, let $\mathcal{U}$ denote the set of user agents and $\mathcal{I}$ denote the set of item agents. 
Each user agent $u \in \mathcal{U}$ maintains a preference memory $\mathcal{M}_u$, which is summarized into a user interest representation $\mathbf{z}_u$. 
Each item agent $i \in \mathcal{I}$ maintains an item-side memory $\mathcal{M}_i$ together with structured item metadata representation $\mathbf{x}_i$.

Given a user agent $u$ and a candidate item set $\mathcal{C}_u = \{c_1, \dots, c_n\}$ produced by an upstream retrieval model, Stage~1 generates a relevance-oriented ranking list by modeling semantic interactions between agents. The user agent evaluates candidates based on their generated semantic descriptions and its preference memory, through a fixed LLM-based inference function:
\begin{equation}
\mathcal{Y}_{u}^{(1)} = f_{\text{LLM}}(u, \mathcal{C}_u),
\end{equation}
where $f_{\text{LLM}}(\cdot)$ denotes the agent interaction module that outputs an ordered list $\mathcal{Y}_{u}^{(1)}$ via semantic reasoning over the full candidate set.

Consistent with AgentCF, the underlying LLM parameters stay frozen throughout: agents adapt by rewriting their memory contents, not by gradient updates. We extend this interaction layer by letting item agents actively generate user-conditioned self-promotion, described next.


\section{Method}
\begin{figure}
\includegraphics[width=\linewidth]{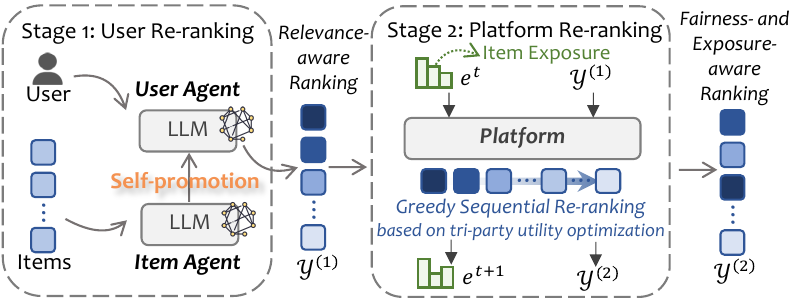}
\caption{Overview of the proposed two-stage TriRec framework,
where Stage 1 performs relevance-aware re-ranking via user–item interaction,
and Stage 2 conducts exposure-aware re-ranking through tri-party utility optimization.
 }

\label{fig:frame}
\end{figure}

Figure~\ref{fig:frame} provides an overview of the proposed TriRec framework.
The framework is composed of two sequential modules: a user--item interaction module for generative personalized self-promotion and a platform control module for exposure-aware re-ranking. 
Responsibility is split deliberately: an item may only write its own self-promotion, while the exposure budget stays with the platform. We do not give items an objective of their own to maximise, because an item that bids for its own exposure invites strategic inflation and leaves the resulting allocation unauditable; keeping the budget on the platform side makes exposure a quantity the operator sets and inspects.
Both modules operate at inference time with frozen LLM parameters: the item agent's memory is initialized in the preference interaction phase of \S\ref{sec:agent_interaction} and then updated online as users arrive (Eq.~\ref{equ:mem_update}), while no gradients are taken at any point. We next describe the design and implementation details of each component.

\subsection{Generative Personalized Self-Promotion}
\label{sec:promotion}

As shown in the left part of Figure~\ref{fig:frame}, Stage~1 models both item agents and user agents, and leverages LLMs to drive their interest expression and interaction behaviors.


Unlike conventional approaches that model items as static feature vectors, we endow item agents with proactive expressive capability. Given the arriving user $u_t \in \mathcal{U}$, the item agent generates a personalized self-promotion as follows:
$$
S_{i \rightarrow u_t}=\mathcal{G}_\theta
\bigl(
\mathbf{x}_i,
\mathbf{z}_{u_t},
\mathcal{M}_i^{t}
\bigr),
$$
where $\mathcal{G}_\theta$ denotes a frozen pre-trained LLM with fixed parameters $\theta$, $\mathbf{x}_i$ represents item metadata, $\mathbf{z}_{u_t}$ denotes the user interest representation, and $\mathcal{M}_i^{t}$ is the memory state of the item agent when $u_t$ arrives.

Self-promotion draws only on the item's own metadata $\mathbf{x}_i$ (title, category, and descriptive features), the target user's interest representation $\mathbf{z}_{u_t}$, and its accumulated memory $\mathcal{M}_i^{t}$; it never accesses test interaction labels, so no test information leaks into the pipeline. The exposure state accumulates across user arrivals in a sequential, causal manner consistent with the online setting. The generation prompt instructs the LLM to ground its output in the provided metadata, which constrains but does not guarantee factuality.

This lets items with very few historical interactions convey their potential value through semantic description rather than accumulated exposure, which is what alleviates the cold-start barrier.

\paragraph{\textbf{Item Memory Update.}}
After serving user $u_t$, the item agent writes the interaction back into its own memory:
\begin{equation}
\mathcal{M}_i^{t+1} =
\mathcal{W}_\theta\bigl(\mathcal{M}_i^{t},\, \mathbf{z}_{u_t},\, S_{i \rightarrow u_t}\bigr),
\label{equ:mem_update}
\end{equation}
where $\mathcal{W}_\theta$ is an LLM-based summarization operator sharing the same frozen parameters $\theta$ as $\mathcal{G}_\theta$. The update consumes only the user representation the promotion was conditioned on and the generated text itself---never the ground-truth interaction or the realized ranking---so no test signal enters memory. Over the arrival stream, $\mathcal{M}_i$ therefore records which kinds of users the item has addressed and how, letting later promotions build on earlier phrasing instead of restating metadata. Because $\mathcal{M}_i^{t}$ conditions only text generation in Stage~1, it does not enter the platform state $\mathbf{s}_t$ of Stage~2.


Instead of relying on an explicit parametric scoring function, the user agent performs semantic preference reasoning over item self-presentations generated by item agents.
Formally, given a candidate item set $\mathcal{C}_u$ and their corresponding self-promotions $\{S_{i \rightarrow u}\}_{i \in \mathcal{C}_u}$, the user agent produces a relevance-oriented ranking by
\begin{equation}
\label{eq:rank_stage1}
\mathcal{Y}_{u}^{(1)} = \arg\!\operatorname{sort}_{i \in \mathcal{C}_u} \;
\mathcal{F}_\phi(\mathbf{z}_u, S_{i \rightarrow u}),
\end{equation}
where $\mathcal{F}_\phi(\cdot)$ denotes the LLM-based semantic preference evaluator with fixed parameters $\phi$. Preference judgments are therefore conditioned on both intrinsic user tastes and context-aware item descriptions.

In practice, since each item agent generates its self-promotion independently conditioned on the target user, all candidates in $\mathcal{C}_u$ can be processed via concurrent API calls, making the wall-clock latency largely independent of the candidate set size.

This agent-based interaction stage is intentionally restricted to relevance-oriented preference alignment: no exposure modulation or fairness constraint is injected here, so the subsequent platform re-ranker regulates exposure on top of a stable relevance backbone.

\subsubsection{LLM Scoring and Calibration}
\label{sec:scoring}
The ranking operator in Eq.~\ref{eq:rank_stage1} is realized by prompting the user
agent to score every candidate on a nonlinear $0$--$10$ scale, and
$\mathcal{Y}_u^{(1)}$ is the descending sort of those scores; the
same scores $r_{\text{LLM}}(u_t,i)$ are consumed by the platform
utility model below. Because generation is stochastic
(temperature $=0.7$), we validate rather than trust the output:
responses are parsed from a strict JSON schema, identifiers outside
$\mathcal{C}_u$ are dropped, duplicates are collapsed to their
highest score, and malformed or incomplete replies are re-queried up
to three times before any remaining candidate is appended to the
list tail, so the scored list always has length $|\mathcal{C}_u|$. Prompt templates are in the released repository.

\subsection{Platform-Led Multi-Objective Re-Ranking}
Although Stage~1 produces relevance-oriented rankings via agent interaction, it does not explicitly regulate long-term exposure allocation across items. 
We therefore introduce a platform re-ranking module $\mathcal{A}_p$ that performs state-aware sequential re-ranking over user utility, item utility, and platform fairness.

We model the platform module as a sequential decision process operating on a dynamic system state.

\paragraph{\textbf{Exposure as a Control State.}}
In real-world systems, the platform cannot directly control user feedback such as clicks or purchases, but can regulate item visibility through ranking positions and display frequency. We therefore adopt exposure as the core system state, characterizing each item's cumulative visibility across recommendation rounds. Exposure also accumulates over time, which left uncontrolled produces long-term popularity bias; modeling it as a dynamic state lets the platform balance immediate utility against exposure accumulated so far in the stream.


\paragraph{\textbf{State Representation.}}
At time step $t$, when user $u_t$ arrives, the platform observes the current system state $\mathbf{s}_t = \{\mathbf{e}^t, \mathcal{Y}_{u_t}^{(1)}\}$,
where $\mathbf{e}^t = [e_i^t]_{i \in \mathcal{I}}$ is the cumulative exposure vector of candidate items $\mathcal{I}$, initialized to zero and updated online as users arrive (Eq.~\ref{equ:update}); $\mathcal{Y}_{u_t}^{(1)}$ is the relevance-oriented ranking list from the Stage~1 agent interaction module. Because $\mathbf{e}^t$ starts from zero while the fairness target $\mathbf{p}_{\text{train}}$ derives from the training distribution, the earliest arrivals receive a stronger fairness push; all results use one fixed arrival order, whose effect we quantify in \S\ref{sec:robustness}.

\paragraph{\textbf{Action Space.}}
Given the observed state $\mathbf{s}_t$, the platform re-ranks the Stage~1 list into the final recommendation $\mathcal{Y}_{u_t}^{(2)} = \mathcal{A}_p(\mathbf{s}_t)$ served to user $u_t$.

\paragraph{\textbf{State Transition.}}
Given the platform re-ranking action $\mathcal{Y}_{u_t}^{(2)} = [i_1, i_2, \dots, i_K]$, 
the exposure state is updated by position-dependent gain:
\begin{equation}
e_i^{t+1} =
\begin{cases}
e_i^t + v(k), & \text{if } i = i_k \in \mathcal{Y}_{u_t}^{(2)}, \\
e_i^t, & \text{otherwise},
\end{cases}
\label{equ:update}
\end{equation}
where $k = \text{rank}_{\mathcal{Y}_{u_t}^{(2)}}(i)$ is the 1-based display position of item $i$,
and $v(\cdot)$ a monotonically decaying exposure function.

In our implementation, we adopt logarithmic decay:
\begin{equation}
v(k) = \frac{1}{\log_2(k+1)},
\label{eq:v}
\end{equation}
so that items placed at higher ranks receive larger exposure increments, 
reflecting position bias~\cite{ndcg_jarvelin}: higher-ranked items attract more attention and interaction. 

Note that $v(\cdot)$ serves a dual role in our 
framework: it defines both the exposure state 
transition (Eq.~\ref{equ:update}) and the 
position-dependent weight $\omega_k$ used in the 
participation policy (Eq.~\ref{equ:alpha_omega}). We compare alternative 
decay profiles in \S\ref{sec:ablation}.

This formulation assumes that only displayed items receive exposure gain, 
while non-displayed items retain their historical exposure levels, and
allows the platform to regulate exposure allocation explicitly.

\paragraph{\textbf{Platform Optimization Objective.}}

Over a sequence of user arrivals, the platform maximizes long-term multi-stakeholder utility by coordinating user relevance, item utility, and fairness.

We decompose the objective into a per-item joint utility function
$U_{\text{joint}}(u_t,i,k,\mathbf{e}^t)$,
which measures the marginal contribution of placing item $i$ at position $k$ for user $u_t$ under exposure state $\mathbf{e}^t$.


Formally, given a time horizon $T$, the objective is defined as:
\begin{equation}
\max_{\{\mathcal{Y}_{u_t}^{(2)}\}_{t=1}^{T}}
\frac{1}{T} \sum_{t=1}^{T}
\sum_{k=1}^{|\mathcal{Y}_{u_t}^{(2)}|}
U_{\text{joint}}(u_t,i,k,\mathbf{e}^t),
\label{max_objection}
\end{equation}
where $\mathcal{Y}_{u_t}^{(2)}$ is the platform re-ranked list served to $u_t$ at step $t$.

\subsubsection{State-Aware Joint Utility Scoring}
\label{sec:method}

Since directly optimizing the induced listwise objective is intractable---the combinatorial search space grows as $O(|\mathcal{C}|! / (|\mathcal{C}|-K)!)$ for a candidate set of size $|\mathcal{C}|$ and list length $K$, making exact enumeration computationally prohibitive---we approximate it using a position-aware marginal joint utility function.

Specifically, we define the position-conditioned joint utility as:
\begin{equation}
U_{\text{joint}}(u_t,i,k,\mathbf{e}^t)
=
g(u_t,i,k,\mathbf{e}^t)
\cdot
U_{\text{expo-item}}(u_t,i,k),
\label{eq:joint_equ}
\end{equation}
where $g(u_t,i,k,\mathbf{e}^t)$ captures the immediate relevance--fairness trade-off at ranking position $k$,
and $U_{\text{expo-item}}(u_t,i,k)$ denotes the exposure-aware item utility modulator.

\paragraph{\textbf{Relevance--Fairness Gain Function.}}

We compute relevance--fairness gain by combining normalized user and platform utility signals through a position-aware convex weighting scheme:
\begin{equation}
g(u_t,i,k,\mathbf{e}^t)
=
\alpha_k \cdot \tilde{U}_{\text{user}}(u_t,i)
+
(1-\alpha_k) \cdot \tilde{U}_{\text{platform}}(i, \mathbf{e}^t),
\end{equation}
where $\alpha_k \in [0,1]$ controls the relative participation of platform-level regulation at rank position $k$, and 
$\tilde{U}_{\text{user}}$ and $\tilde{U}_{\text{platform}}$ denote the normalized versions of the corresponding utility signals, rescaled to the same range for stable convex combination.

For numerical stability and scale consistency, all heterogeneous utility signals are normalized within the candidate set at each time step. We use $\tilde{U}_{(\cdot)}$ to denote the normalized versions of the corresponding raw utility signals.

\paragraph{\textbf{User Utility Modeling.}}

The user-side utility is the preference intensity predicted by the user agent through semantic reasoning over the full candidate set, $U_{\text{user}}(u_t,i) = r_{\text{LLM}}(u_t,i)$, normalized across candidates to obtain $\tilde{U}_{\text{user}}$.

\paragraph{\textbf{Platform Utility Modeling.}}

The platform utility term aggregates normalized marginal fairness gains:
\begin{equation}
\tilde{U}_{\text{platform}}(i, \mathbf{e}^t)
=
\lambda_1 \cdot \tilde{U}_{\text{DGU}}(i, \mathbf{e}^t)
+
\lambda_2 \cdot \tilde{U}_{\text{MGU}}(i, \mathbf{e}^t),
\end{equation}
where $\tilde{U}_{\text{DGU}}$ and $\tilde{U}_{\text{MGU}}$ denote normalized marginal improvements on Distributional Group Unfairness and Maximal Group Unfairness, respectively. 
Specifically, for each candidate item $i$, we compute the marginal fairness gain by simulating its placement at the current position: the exposure distribution is tentatively updated as $\mathbf{e}^{t'} = \mathbf{e}^t + v(k) \cdot \mathbf{1}_i$, and the resulting reduction in DGU (or MGU) relative to the current state is taken as the marginal improvement, i.e., $U_{\text{DGU}}(i, \mathbf{e}^t) = \text{DGU}(\mathbf{e}^t) - \text{DGU}(\mathbf{e}^{t'})$.
The coefficients $\lambda_1$ and $\lambda_2$ control the relative importance of the two fairness objectives.

\paragraph{\textbf{Position-Aware Participation Policy.}}

To regulate intervention strength along the list, we use a monotonic position-dependent policy:
\begin{equation}
\alpha_k
=
\alpha_{\min}
+
(\alpha_{\max} - \alpha_{\min})
\cdot
(\omega_k)^p,
\label{equ:alpha_omega}
\end{equation}
where $p$ controls the decay curvature, $\alpha_{\min}$ and $\alpha_{\max}$ define the lower and upper bounds, and $\omega_k \in [0,1]$ is a position-dependent weight derived from the same exposure decay profile $v(\cdot)$ used in the state transition function Eq.~\ref{eq:v}, with larger values corresponding to higher-ranked positions.

This design prioritizes user relevance at top-ranked positions while gradually injecting platform-level regulation toward lower ranks, enabling a controllable relevance--fairness trade-off.

\paragraph{\textbf{Exposure-Aware Item Utility.}}
Given that the ranking position is explicitly determined during sequential construction,
we rewrite item-side utility defined in Eq.~\ref{uitem} as:
\begin{equation}
U_{\text{item}}(u_t,i,k)
=
v(k) \cdot \text{CTR}(u_t,i),
\end{equation}
where $v(k)$ models position-dependent exposure probability and $\text{CTR}(u_t,i)$ denotes the predicted click-through probability:
\begin{equation}
    \text{CTR}(u_t,i)
    = \sigma\!\Big(\mathrm{sim}_{\mathrm{emb}}(\mathbf{z}_{u_t}, 
    \mathbf{h}_i)\Big),
\end{equation}
where $\sigma(\cdot)$ is the sigmoid function, $\mathrm{sim}_{\text{emb}}(\cdot)$ represents cosine similarity between the user interest embedding $\mathbf{z}_{u_t}$ and the item semantic embedding $\mathbf{h}_i$. 

To incorporate long-term exposure regulation, we define the exposure-aware item utility as
\begin{equation}
U_{\text{expo-item}}(u_t,i,k)
=
\big(\tilde{U}_{\text{item}}(u_t,i,k)\big)^{\lambda_{\text{item}}},
\end{equation}
where $\tilde{U}_{\text{item}}$ denotes item utility within the candidate set at each ranking step, 
and $\lambda_{\text{item}}$ controls the re-ranker's sensitivity to item exposure utility.

This lets the platform favor items with high expected click value at each step; under-exposure is handled separately by the group-fairness term above.

\subsubsection{Greedy Sequential Action Generation}

In implementation, Eq.~\ref{max_objection} is approximated via greedy list construction performed independently for each incoming user request, 
while the exposure state $\mathbf{e}^t$ is carried over across recommendation rounds, 
effectively realizing a one-step approximation of the long-term control objective.
We note that this myopic approximation does not guarantee globally optimal long-term behavior; however, carrying persistent exposure state across rounds provides implicit temporal coordination that empirically yields strong multi-stakeholder performance (see \S\ref{sec:experiments}).

Due to the position-dependent nature of $U_{\text{joint}}(u_t,i,k,\mathbf{e}^t)$, 
the final ranking list is constructed in a top-down manner, 
where items are greedily selected and fixed from higher to lower ranks.

At ranking position $k$, the platform re-ranker selects:
\begin{equation}
i_k^* = \arg\max_{i \in \mathcal{C}_{u_t} \setminus \mathcal{Y}_{<k}} 
U_{\text{joint}}(u_t,i,k,\mathbf{e}^t),
\end{equation}
where $\mathcal{Y}_{<k} = \{i_1^*, \dots, i_{k-1}^*\}$ denotes the prefix list containing the previously selected top-$k-1$ items.

This sequential construction explicitly accounts for position-aware utility modulation and enables efficient approximation of the underlying listwise optimization objective. This greedy sequential construction thus serves as the concrete policy implementation of the platform re-ranking module, mapping observed states to ranking actions under the defined utility objective.

Although greedy construction is not globally optimal, it is widely adopted in practical re-ranking systems due to its efficiency and strong empirical performance. Notably, this re-ranking step involves only closed-form arithmetic operations over the candidate set, making it computationally negligible.

\subsubsection{Platform Re-Ranking as a Closed-Loop Decision Process}

Finally, the platform module is a sequential control process $\mathcal{A}_p = \langle \mathcal{S}, \mathcal{A}, \mathcal{T} \rangle$,
where the state space $\mathcal{S}$ consists of the historical exposure vector $\mathbf{e}^t$ and the Stage~1 relevance ranking $\mathcal{Y}_{u_t}^{(1)}$; 
the action space $\mathcal{A}$ corresponds to the re-ranking decision $\mathcal{Y}_{u_t}^{(2)}$; 
and the transition function $\mathcal{T}$ is defined by the exposure update rule in Eq.~\ref{equ:update}.

\section{Experiments}
\label{sec:experiment}
\label{sec:experiments}


We address the following research questions:
\begin{itemize}[leftmargin=*]
    \item \textbf{RQ1:} How does TriRec perform in accuracy, fairness, and expected item utility against existing methods?
    \item \textbf{RQ2:} What do individual components and design 
          choices contribute to the final performance?
    \item \textbf{RQ3:} How do key hyperparameters affect the tri-party trade-off?
    \item \textbf{RQ4:} How effective is self-promotion at improving the ranking of cold-start items?
\end{itemize}

\subsection{Experimental Setup}

\subsubsection{Dataset}
We conduct experiments on four real-world datasets from three sources: \textbf{Amazon}\footnote{\url{https://amazon-reviews-2023.github.io/}} (CDs \& Vinyl, Movies \& TV), \textbf{Steam}\footnote{\url{https://cseweb.ucsd.edu/~jmcauley/datasets.html\#steam_data}} (Games), and \textbf{Goodreads}\footnote{\url{https://mengtingwan.github.io/data/goodreads.html}} (Young Adult). Following standard practice, we keep positive interactions (rating $\geq$ 4) and users with 10--100 interactions, subsample each dataset to $2{,}000$ users preserving the interaction distribution (Table~\ref{tab:dataset_stats}), and encode items with \texttt{text-embedding-ada-002}.

\begin{table}[h]
\centering
\caption{Statistics of the evaluation subsets ($2{,}000$ users each).}
\label{tab:dataset_stats}
\begin{tabular}{lrrr}
\toprule
\textbf{Datasets} & \textbf{\#Items} & \textbf{\#Inters.} & \textbf{Sparsity} \\ \midrule
CDs \& Vinyl & 33,042 & 41,901 & 99.93\% \\
Movies \& TV & 27,407 & 40,907 & 99.92\% \\
Goodreads YA & 10,786 & 60,743 & 99.71\% \\
Steam Games & 6,635 & 36,006 & 99.72\% \\ \bottomrule
\end{tabular}
\end{table}

\begin{table*}[t]
\centering
\caption{Overall performance comparison across four datasets. 
The best results are bolded.}
\label{tab:overall_performance}
\setlength{\tabcolsep}{1.5pt} 
\resizebox{\textwidth}{!}{
\begin{tabular}{l ccccc c ccccc c ccccc c ccccc}
\toprule
\multirow{2}{*}{Method} & \multicolumn{5}{c}{CDs \& Vinyl} && \multicolumn{5}{c}{Movies \& TV} && \multicolumn{5}{c}{Goodreads YA} && \multicolumn{5}{c}{Steam Games} \\
\cmidrule{2-6} \cmidrule{8-12} \cmidrule{14-18} \cmidrule{20-24}
& NDCG$\uparrow$ & MRR$\uparrow$ & DGU$\downarrow$ & MGU$\downarrow$ & EIU$\uparrow$ && NDCG$\uparrow$ & MRR$\uparrow$ & DGU$\downarrow$ & MGU$\downarrow$ & EIU$\uparrow$ && NDCG$\uparrow$ & MRR$\uparrow$ & DGU$\downarrow$ & MGU$\downarrow$ & EIU$\uparrow$ && NDCG$\uparrow$ & MRR$\uparrow$ & DGU$\downarrow$ & MGU$\downarrow$ & EIU$\uparrow$ \\
\midrule
MACRec & 0.4858 & 0.4676 & 0.1632 & 0.1467 & 0.5898 &&
    0.4167 & 0.4032 & 0.2322 & 0.1787 & 0.5375 &&
    0.5431 & 0.5200 & 0.5568 & 0.5323 & 0.6279 &&
    0.3327 & 0.3367 & 0.4483 & 0.4314 & 0.4857 \\
AgentCF++ & 0.3874 & 0.3914 & 0.1587 & 0.1479 & 0.5289 &&
    0.3980 & 0.3960 & 0.2275 & 0.1797 & 0.5314 && 
    0.5004 & 0.4816 & 0.5336 & \textbf{0.4942} & 0.5990 &&
    0.4431 & 0.4238 & 0.4188 & 0.4023 & 0.5547 \\
Rec4AgentVerse & 0.3878 & 0.3546 & 0.1705 & 0.1588 & 0.4998 &&
    0.4476 & 0.4168 & 0.2479 & 0.1970 & 0.5528 &&
    0.5354 & 0.4881 & 0.6389 & 0.6325 & 0.5923 &&
    0.4088 & 0.3702 & 0.4984 & 0.4747 & 0.5139 \\
DualRec & 0.3130 & 0.3063 & 0.1572 & 0.1549 & 0.4637 &&
    0.3097 & 0.3081 & 0.2504 & 0.1961 & 0.4645 &&
    0.2509 & 0.2619 & 0.6481 & 0.6301 & 0.4268 &&
    0.2949 & 0.2935 & 0.5038 & 0.4798 & 0.4530 \\
SCRUF-D & 0.2580 & 0.2681 & \textbf{0.1508} & 0.1465 & 0.4340 &&
    0.3078 & 0.2968 & 0.2394 & 0.1898 & 0.4564 &&
    0.2645 & 0.2736 & 0.6387 & 0.6315 & 0.4371 &&
    0.3229 & 0.3124 & 0.4914 & 0.4672 & 0.4684 \\
LTP-MMF & 0.4656 & 0.4580 & 0.1780 & 0.1700 & 0.5737 &&
     0.4354 & 0.4519 & 0.3220 & 0.2286 & 0.5628 &&  
     0.5367 & 0.5180 & 0.5204 & 0.5127 & 0.6223 &&  
     0.4441 & \textbf{0.4561} & 0.4342 & 0.4288 & 0.5684 \\
\midrule
\textbf{TriRec} & \textbf{0.4988} & \textbf{0.4728} & 0.1589 & \textbf{0.1462} & \textbf{0.5946} &&
    \textbf{0.5009} & \textbf{0.4664} & \textbf{0.2259} & \textbf{0.1766} & \textbf{0.6089} && 
    \textbf{0.6351} & \textbf{0.6033} & \textbf{0.5073} & 0.5021 & \textbf{0.6939} &&
    \textbf{0.4767} & 0.4461 & \textbf{0.3591} & \textbf{0.3588} & \textbf{0.5874} \\
\bottomrule
\end{tabular}}
\end{table*}

\subsubsection{Baselines}
We compare TriRec against three groups of baselines:
\begin{itemize}[leftmargin=*]
    \item \textbf{Agent-based User-Centric:} 
    \textit{AgentCF++}~\cite{agentcf++} (memory-enhanced 
    LLM-based agent interactions; we adopt its 
    single-domain configuration) and 
    \textit{MACRec}~\cite{macrec} (multi-agent 
    collaboration). Both optimize only for user relevance.
    \item \textbf{Creator-Side Focus:} 
    \textit{Rec4Agentverse}~\cite{rec4agentverse} (item 
    agents for autonomous interaction) and 
    \textit{DualRec}~\cite{dualrec} (items as active 
    queries in a dual-market setting). Both model 
    item-side dynamics but without explicit exposure 
    optimization or self-promotion.
    \item \textbf{Platform Fairness Re-ranking:} 
    \textit{SCRUF-D}~\cite{scruf_d} (multi-agent social 
    choice for fairness arbitration) and 
    \textit{LTP-MMF}~\cite{ltp_mmf}, the long-term 
    extension of P-MMF~\cite{pmmf}, which enforces provider 
    max-min fairness under recommendation feedback loops. Both inject 
    fairness at the platform level via post-hoc re-ranking.
\end{itemize}

All LLM-based baselines (AgentCF++, Rec4AgentVerse, MACRec)
run on the same \texttt{gpt-4o-mini} backbone with the same item memory and
user description as TriRec, so none benefits from a stronger LLM or prompt.
SCRUF-D, LTP-MMF, and DualRec use no LLM, forming a deterministic non-LLM
provider-fairness control against which the Stage-2 policy is judged.

\subsubsection{Evaluation Protocol}

We evaluate from three stakeholder perspectives:
\textbf{(1) User accuracy:} NDCG@K (K=5) and MRR, measuring 
ranking quality against ground-truth interactions.
\textbf{(2) Item utility:} Expected Item Utility (EIU, 
Eq.~\ref{uitem}), the position-weighted click propensity 
aggregated across users.
\textbf{(3) Platform fairness:} DGU@K and 
MGU@K~\cite{flower} (K=10), measuring distributional and 
worst-case group exposure disparity over eight 
popularity groups.

\subsubsection{Candidate Construction.}
We adopt a leave-one-out protocol, using each user's 
most recent interaction as the test item and the 
remainder for training. Each candidate set contains 
1 ground-truth item and $n$ negatives. 
Unlike prior work~\cite{agentcf} that samples negatives 
uniformly at random, we retrieve \emph{hard} negatives 
with a pre-trained SASRec model~\cite{sasrec}, so every 
candidate has already passed an upstream retrieval filter 
and sits near the decision boundary rather than being 
trivially separable. Crucially, \textbf{all 
methods are evaluated on identical candidate sets}, so 
differences are attributable to the algorithms rather 
than to candidate construction.

We evaluate under candidate pools from 10 to 50
items. The base $|\mathcal{C}_u|{=}10$ ($n{=}9$) follows the established
LLM-agent protocol~\cite{agentcf, agentcf++} for like-for-like comparison;
$|\mathcal{C}_u|\in\{10,15,20\}$ tests robustness to discrimination difficulty
(\S\ref{sec:robustness}); and the mechanism ablation runs at $|\mathcal{C}_u|{=}50$
(\S\ref{sec:coldstart}), the upper end of the 10--50 range industrial
final-stage re-rankers operate on~\cite{covington2016deep, huang2013learning}.
Across $2{,}000$ users this is $2{\times}10^{4}$ ranking decisions per dataset.

\subsubsection{Platform Re-Ranking Configuration.}
We use \texttt{gpt-4o-mini} as the default backbone for all agents (alternative
backbones in \S\ref{sec:backbone}). The position-aware participation policy uses
$\alpha_{\max}{=}1.0$, $\alpha_{\min}{=}0.1$, $p{=}0.1$; fairness coefficients
$\lambda_1{=}\lambda_2{=}0.5$; and item-utility exponent $\lambda_{\text{item}}{=}10$.
Sensitivity analysis of all hyperparameters is 
provided in \S\ref{sec:rq3}.

\subsection{Overall Performance Comparison (RQ1)}
Table~\ref{tab:overall_performance} reports the overall 
comparison across four datasets, analyzed from three 
stakeholder perspectives.

\subsubsection{User-Side Accuracy.}
TriRec achieves the highest NDCG across all four datasets
and the highest MRR on three of them; on Steam Games,
LTP-MMF's slightly higher MRR comes from aggressively promoting the single
top item at a large fairness cost (below). The gains over user-centric agents
(MACRec, AgentCF++) and creator-side methods (Rec4AgentVerse, DualRec)
indicate that active, personalized item expression beats passive modeling.

To quantify reliability, we compute per-user bootstrap $95\%$ confidence
intervals ($B{=}2000$) and paired tests of TriRec against the strongest
baseline on each dataset (NDCG@5). The gain is statistically significant on
three of the four datasets: Goodreads YA ($\Delta{=}{+}0.092$ vs.\ MACRec,
$p{<}0.001$), Movies \& TV (${+}0.053$ vs.\ Rec4AgentVerse, $p{<}0.001$), and
Steam Games (${+}0.033$ vs.\ LTP-MMF, $p{=}0.004$). On CDs
\& Vinyl the margin over MACRec is positive but within noise (${+}0.013$,
$p{=}0.125$), so we claim a reliable accuracy advantage on three datasets
rather than uniformly.

\subsubsection{Platform-Level Fairness.}
TriRec attains the lowest DGU on Movies \& TV, Goodreads YA, and Steam Games,
and the lowest MGU on CDs \& Vinyl, Movies \& TV, and Steam Games. Its CDs DGU
($0.159$) is mid-pack---SCRUF-D ($0.151$), DualRec ($0.157$) and AgentCF++
($0.159$) are lower---but each buys that parity with far less relevance (NDCG
$0.258$--$0.387$ vs.\ $0.499$); on Goodreads YA AgentCF++ reaches a lower MGU
($0.494$ vs.\ $0.502$) at markedly weaker accuracy and item utility (NDCG
$0.500$, EIU $0.599$). LTP-MMF, despite strong accuracy, has the worst fairness
on both Amazon datasets, showing that optimizing engagement feedback alone
cannot balance group exposure without explicit regulation.

\subsubsection{Item-Side Utility.}
\label{sec:itemutility}
TriRec achieves the highest EIU on all four datasets, from the combined effect of
item self-promotion (semantic signals that improve matching) and exposure-aware
re-ranking (shifting exposure to under-served items with high click probability).
The margin over creator-side methods (Rec4AgentVerse, DualRec), which also model
item-side dynamics, shows active self-expression yields greater item-side benefit
than passive dual-market modeling.

Although Stage~2 optimizes the same quantity that EIU reports, we validate the
underlying CTR proxy against held-out interactions: ranking each user's candidates
by this proxy places the held-out ground-truth item above a randomly drawn
non-interacted candidate $70.3\%$ of the time (within-user AUC $0.7025$,
CI $[0.690,0.715]$; $N{=}1{,}999$, CDs \& Vinyl, $|\mathcal{C}_u|{=}50$).
Crucially, this validity is not inherited from
popularity: on the $1{,}371$ users whose ground-truth item has \emph{no}
training interactions, the proxy still attains within-user AUC $0.7012$,
so its discriminative power must come from semantic matching rather than
interaction history.
EIU nonetheless remains a term in the objective TriRec optimizes, so we 
also report an independent tail measure: the re-ranker's net 
contribution to the position-weighted share of top-$K$ 
exposure reaching zero-exposure items, taken relative to the 
candidate pool's own composition. That contribution is 
$+1.40$pp on CDs \& Vinyl and $+1.28$pp on Movies \& TV, but 
only $+0.10$pp on Goodreads YA and $-0.16$pp on Steam Games. 
Re-ranking promotes the tail where the retrieved pool actually 
contains zero-exposure items, and cannot manufacture tail 
exposure where it does not.

\subsubsection{Generalization and Robustness.}
\label{sec:robustness}
The four datasets span e-commerce (CDs \& Vinyl,
Movies \& TV), social reading (Goodreads YA), and gaming
(Steam); TriRec's relative advantages remain stable across them.
Varying the pool $|\mathcal{C}_u|\in\{10,15,20\}$ on CDs \& Vinyl leaves
the ordering intact---all methods degrade as negatives grow more
competitive, but TriRec stays best or near-best on all three metrics.
Arrival order is a further confound, since the exposure state accumulates
from zero; replaying Stage~2 under $20$ random shuffles
(CDs \& Vinyl, $|\mathcal{C}_u|{=}50$, \textbf{full}) leaves every metric
fixed---NDCG@10 $0.2547{\pm}0.0001$, DGU@10 $0.1270{\pm}0.0006$, MGU@10
$0.1065{\pm}0.0002$, zero-exposure share $8.158\%{\pm}0.005$---so the
conclusions do not hinge on a particular ordering.

\begin{table}[t]
\centering
\caption{Ablation study on CDs \& Vinyl.}
\label{tab:ablation}
\setlength{\tabcolsep}{1.5pt}
\begin{tabular}{lccccc}
\toprule
\textbf{Model Variants} 
& $\mathbf{NDCG}\uparrow$ 
& $\mathbf{MRR}\uparrow$ 
& $\mathbf{DGU}\downarrow$ 
& $\mathbf{MGU}\downarrow$ 
& $\mathbf{EIU}\uparrow$ \\
\midrule

(a) w/o $\mathcal{Y}^{(1)}$ 
& 0.3381 & 0.3240 & 0.1298 & 0.1309 & 0.4782 \\

(b) w/o \{$S_{i \rightarrow u}$, $\mathcal{Y}^{(2)}$\} 
& 0.3642 & 0.3718 & 0.1695 & 0.1538 & 0.5143 \\

(c) w/o $\mathcal{Y}^{(2)}$ 
& 0.4684 & 0.4581 & 0.1659 & 0.1477 & 0.5816 \\

\midrule

(d) w/o $U_{\text{platform}}$ 
& 0.4873 & 0.4615 & 0.1683 & 0.1556 & 0.5851 \\

(e) w/o Dynamic $\alpha_k$ 
& 0.4635 & 0.4276 & 0.1301 & 0.1296 & 0.5603 \\

(f) w/o $U_{\text{user}}$ 
& 0.2634 & 0.2839 & 0.0815 & 0.0823 & 0.4458 \\

(g) w/o $U_{\text{item}}$ 
& 0.4857 & 0.4579 & 0.1535 & 0.1460 & 0.5824 \\

\midrule

(h) w/o $\mathrm{Sim}_{\mathrm{emb}}(\mathbf{z}, \mathbf{h})$ 
& 0.4776 & 0.4487 & 0.1560 & 0.1470 & 0.5753 \\

(i) with $\mathrm{Sim}^{\mathrm{rand}}_{\mathrm{emb}}(\mathbf{z}, \mathbf{h})$ 
& 0.4627 & 0.4283 & 0.1653 & 0.1532 & 0.5601 \\

\midrule

\textbf{TriRec} 
& \textbf{0.4988} & \textbf{0.4728} & 0.1589 & 0.1462 & \textbf{0.5946} \\

\bottomrule
\end{tabular}
\end{table}

\subsection{Ablation Study (RQ2)}
\label{sec:ablation}
Table~\ref{tab:ablation} summarizes ablation results on CDs \& Vinyl.

\subsubsection{Stage-Wise Component Analysis.}
Removing Stage~1 (row~(a)) drops NDCG by $31.7\%$, confirming LLM-driven
semantic interaction is indispensable. Comparing rows~(b) and~(c) isolates item
self-promotion, a $10.4$-point NDCG gain; as these rows differ in more than one
factor, we defer controlled attribution to Section~\ref{sec:coldstart}. Removing
Stage~2 (row~(c)) keeps accuracy but degrades fairness (DGU $0.170$) and item
utility (EIU $0.582$), confirming platform-level regulation is necessary.

\begin{table*}[!t]
\centering
\caption{Impact of different position decay functions on recommendation performance.}
\label{tab:decay_functions}
\setlength{\tabcolsep}{1.5pt}
\resizebox{\textwidth}{!}{
\begin{tabular}{l ccccc c ccccc c ccccc c ccccc}
\toprule
\multirow{2}{*}{Decay Function} & \multicolumn{5}{c}{CDs \& Vinyl} && \multicolumn{5}{c}{Movies \& TV} && \multicolumn{5}{c}{Goodreads YA} && \multicolumn{5}{c}{Steam Games} \\
\cmidrule{2-6} \cmidrule{8-12} \cmidrule{14-18} \cmidrule{20-24}
& NDCG$\uparrow$ & MRR$\uparrow$ & DGU$\downarrow$ & MGU$\downarrow$ & EIU$\uparrow$ 
&& NDCG$\uparrow$ & MRR$\uparrow$ & DGU$\downarrow$ & MGU$\downarrow$ & EIU$\uparrow$
&& NDCG$\uparrow$ & MRR$\uparrow$ & DGU$\downarrow$ & MGU$\downarrow$ & EIU$\uparrow$
&& NDCG$\uparrow$ & MRR$\uparrow$ & DGU$\downarrow$ & MGU$\downarrow$ & EIU$\uparrow$ \\
\midrule

Power-law $\frac{1}{k^q}$ & 0.4948 & 0.4716 & 0.1594 & 0.1468 & 0.5934
&& 0.4985 & 0.4632 & 0.2355 & 0.1774 & 0.6068  
&& \textbf{0.6353} & \textbf{0.6035} & 0.5075 & 0.5023 & \textbf{0.6941}
&& 0.4758 & 0.4459 & 0.3597 & 0.3585 & 0.5872 \\

Exponential $e^{-\lambda(k-1)}$ & 0.4937 & 0.4709 & 0.1597 & 0.1472 & 0.5928 
&& 0.4991 & 0.4638 & 0.2359 & 0.1777 & 0.6072
&& 0.6351 & 0.6034 & 0.5074 & \textbf{0.5020} & 0.6940
&& 0.4755 & 0.4441 & 0.3613 & \textbf{0.3584} & 0.5859 \\

Linear $1-\frac{k-1}{K}$ & \textbf{0.4998} & \textbf{0.4729} & 0.1610 & 0.1472 & \textbf{0.5947}
&& 0.4984 & 0.4634 & 0.2364 & 0.1779 & 0.6069
&& 0.6346 & 0.6031 & 0.5075 & 0.5024 & 0.6937
&& 0.4753 & 0.4455 & 0.3597 & 0.3593 & 0.5868 \\

Log $\frac{1}{\log_2(k+1)}$ & 0.4988 & 0.4728 & \textbf{0.1589} & \textbf{0.1462} & 0.5946 && \textbf{0.5009} & \textbf{0.4664} & \textbf{0.2259} & \textbf{0.1766} & \textbf{0.6089} 
&& 0.6351 & 0.6033 & \textbf{0.5073} & 0.5021 & 0.6939
&& \textbf{0.4767} & \textbf{0.4461} & \textbf{0.3591} & 0.3588 & \textbf{0.5874}\\

\bottomrule
\end{tabular}}
\end{table*}

\subsubsection{Platform-Level and Representation Analysis.}
Removing $U_{\text{platform}}$ (row~(d)) keeps accuracy but limits fairness;
fixing $\alpha_k$ instead of decaying it (row~(e)) over-optimizes top-position
fairness and drops NDCG $0.499{\to}0.464$. Removing $U_{\text{user}}$ (row~(f))
causes the sharpest accuracy loss (NDCG $0.263$), and removing
$U_{\text{item}}$ (row~(g)) hurts both accuracy and EIU, confirming that item
utility and user satisfaction are positively coupled. Finally, dropping
semantic embeddings (rows~(h),(i)) degrades both (NDCG $0.478{\to}0.463$),
showing they provide essential grounding.

\subsubsection{Effect of Position Decay Function $v(\cdot)$.}
Table~\ref{tab:decay_functions} compares four monotonic decay forms---Log
$1/\log_2(k{+}1)$ (default), Power-law $1/k^q$, Exponential
$e^{-\lambda(k-1)}$, and Linear $1{-}(k{-}1)/K$ ($q{=}1$, $\lambda{=}0.5$,
$K{=}10$)---which changes NDCG by at most $1.3\%$ on all four datasets
($0.494$--$0.500$ on CDs \& Vinyl); we adopt Log for its most consistent
fairness advantage.

\subsubsection{Effect of LLM Backbone}      
\label{sec:backbone}

\begin{table}[!t]
\centering
\caption{TriRec with different LLM backbones.}
\label{tab:backbone}
\setlength{\tabcolsep}{4pt}
\begin{tabular}{lccccc}
\toprule
Backbone &NDCG$\uparrow$ & MRR$\uparrow$ & DGU$\downarrow$ & MGU$\downarrow$ & EIU$\uparrow$ \\
\midrule
GPT-3.5-Turbo &0.4964 & 0.4718 &0.1568 & 0.1478& 0.5937\\
GPT-4o-mini & 0.4988 & 0.4728 & 0.1589 & 0.1462 & 0.5946\\
GPT-4o & 0.5234& 0.4901& 0.1647 &0.1481 & 0.6086\\
DeepSeek-V3 &0.5075 & 0.4770&0.1635 & 0.1526&0.5986 \\
Qwen-plus &0.4891 & 0.4621 &0.1604 & 0.1521& 0.5864\\
\bottomrule
\end{tabular}
\end{table}
Table~\ref{tab:backbone} reports TriRec with five LLM backbones on CDs \& Vinyl. All yield closely matched NDCG and EIU---GPT-3.5-Turbo to GPT-4o span only a small relevance--fairness trade-off, and DeepSeek-V3 and Qwen behave similarly---so the gains stem from the tri-party design, not a specific LLM.

\subsubsection{Efficiency.}
Following~\cite{moa}, we report average LLM tokens per user request as a hardware-agnostic proxy. On CDs \& Vinyl, TriRec consumes 98.4 tokens/user, comparable to MACRec (114.3) and well below Rec4AgentVerse (143.7); AgentCF++ uses only 41.8 tokens but with substantially weaker accuracy and item utility. In wall-clock terms (\texttt{gpt-4o-mini}, 36-way concurrency), the stages differ by three orders of magnitude: Stage~2 re-ranks a user in $9.5$ms of local arithmetic, whereas Stage~1 generation dominates and grows with the pool ($1.4$/$17.5$s per user at $|\mathcal{C}_u|{=}10$/$50$), making TriRec an offline/batch re-ranker at the larger pool. This cost is not intrinsic: a promotion depends on item metadata and the user's stored interest representation, both known before the request, so pitches can be precomputed in batch, leaving only the $9.5$ms Stage~2 online. The cache then grows with (user, candidate) pairs; dropping the user conditioning shrinks it to one pitch per item, at the price of exposure but not accuracy (\S\ref{sec:coldstart}).

\begin{figure}[t]
    \centering
    \includegraphics[width=\linewidth]{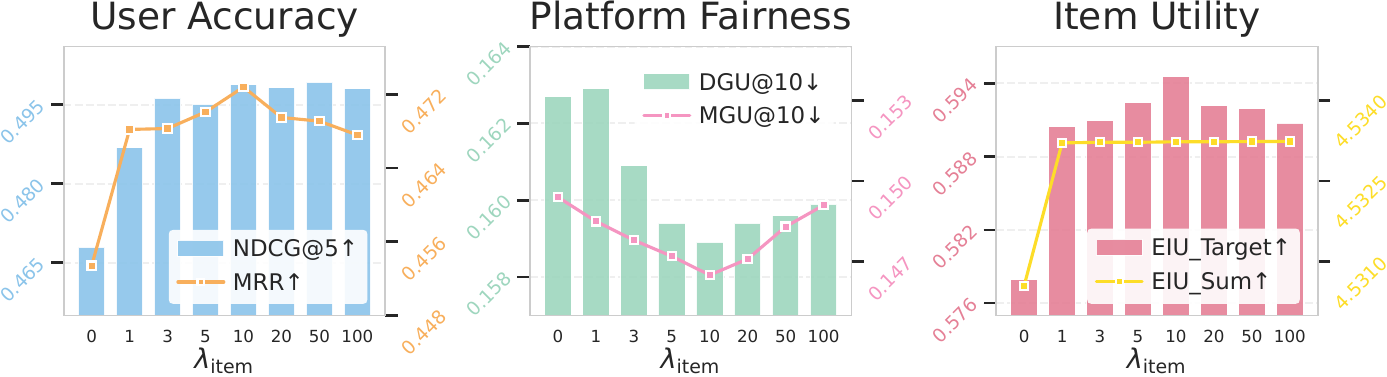}
    \caption{Performance with varying $\lambda_{\text{item}}$ on the CDs dataset.}
    \label{fig:lambda_item}
\end{figure}



\subsection{Analysis of Key Factors (RQ3)}
\label{sec:rq3}
We run sensitivity analysis on CDs \& Vinyl, with $\lambda_{\text{item}}$ as the
primary factor, followed by $\alpha_{\max}$, $p$, $\lambda_1$, and $\lambda_2$.

\subsubsection{Effect of $\lambda_{\text{item}}$.}
Figure~\ref{fig:lambda_item} varies $\lambda_{\text{item}}$ from 0 to 100.
At $\lambda_{\text{item}}{=}0$ the modulator degrades to a constant and all
three stakeholder metrics are worst, confirming item-utility modeling is
essential. Raising it to $5$--$10$ improves all three at once (NDCG to
$0.499$, DGU to $0.159$, EIU to $0.595$)---challenging the assumption that
stakeholder objectives must conflict, as the modulator surfaces
under-exposed items that benefit users, items, and fairness together.
Beyond $20$ performance plateaus and cumulative EIU stays constant
($\approx4.533$), so $\lambda_{\text{item}}$ redistributes exposure rather than
adding volume. We set $\lambda_{\text{item}}{=}10$.

\subsubsection{Effect of Other Hyperparameters.}
The remaining parameters show no sensitivity that changes our conclusions.
$\alpha_{\max}$ trades accuracy for fairness (raising it $0.1{\to}1.0$ moves
NDCG $0.434{\to}0.499$ and DGU $0.105{\to}0.159$); we set $\alpha_{\max}{=}1.0$
because the sparse exposure history here makes aggressive top-position fairness
injection counterproductive, while lower ranks still retain
$\alpha_k{\approx}0.1$ so regulation is not disabled. Larger $p$ shifts the
same trade-off toward fairness (DGU $0.159{\to}0.140$, NDCG $0.499{\to}0.482$),
with accuracy decaying fast beyond $p{=}1.0$, so we keep $p{<}1.0$.
$\lambda_1,\lambda_2$ are robust: varying either $0{\to}1$ improves its fairness
metric at ${<}0.3\%$ NDCG loss.

\subsection{Cold-Start Item Promotion (RQ4)}
\label{sec:coldstart}

\subsubsection{Decomposing the Item-Side Effect.}
Agentic recommenders report end-to-end gains without separating how much comes
from supplying a written promotion at all versus conditioning that promotion on
the target user; the two are routinely bundled. Rows~(b) and~(c) of
Table~\ref{tab:ablation} differ in more than one factor and cannot tell us
\emph{why} self-promotion helps, so we run a controlled ablation on
CDs \& Vinyl ($|\mathcal{C}_u|{=}50$) whose
three arms differ only in the item-side input and share
\emph{identical} candidate sets (verified per user):
\textbf{none} feeds the ranker raw metadata, \textbf{generic}
an LLM-written promotion blind to the target user---a plain LLM
rewrite of that same metadata---and
\textbf{full} the user-conditioned self-promotion of
Section~\ref{sec:promotion}. Thus \textbf{generic} versus
\textbf{none} isolates \emph{expression}, and \textbf{full}
versus \textbf{generic} isolates \emph{conditioning} while
holding LLM rewriting fixed.

Table~\ref{tab:promo_decomp} reports the decomposition. For
items with zero training exposure, expression alone lifts
NDCG@10 from $0.190$ to $0.242$ ($\Delta{=}{+}0.052$,
$p<0.001$), whereas conditioning adds a further $+0.012$
that is not significant ($p{=}0.165$; the $95\%$ CI upper
bound of $+0.027$ bounds any residual accuracy benefit). The
item side behaves differently: the position-weighted share of
top-$10$ exposure going to zero-exposure items rises from
$5.68\%$ to $7.62\%$ under expression and to $8.16\%$ under
conditioning---a $43.6\%$ relative increase over
\textbf{none}, both increments significant. Conditioning
therefore serves the item-side objective of securing exposure
rather than the platform's accuracy objective, consistent
with our framing of items as parties whose interest is
visibility. The effect is also specific to cold items: for
targets with $1$--$5$ training interactions the three arms
are statistically indistinguishable ($0.259$, $0.261$,
$0.254$).

\begin{table}[t]
\centering
\caption{Significance-tested decomposition of the self-promotion effect on
\emph{cold} (zero-exposure) target items, at $5{\times}$ the default candidate
pool (CDs \& Vinyl, $|\mathcal{C}_u|{=}50$, identical candidate sets across
arms, $N{\approx}1{,}370$). Rows isolate expression (none$\to$generic),
conditioning (generic$\to$full), and their total; $\Delta$ is a per-user paired
difference and $p$ a paired permutation test.}
\label{tab:promo_decomp}
\setlength{\tabcolsep}{4pt}
\begin{tabular}{@{}l rr rr@{}}
\toprule
& \multicolumn{2}{c}{\textbf{NDCG@10}} & \multicolumn{2}{c}{\textbf{Exposure share}} \\
\cmidrule(lr){2-3} \cmidrule(lr){4-5}
\textbf{Mechanism} & $\Delta$ & $p$ & $\Delta$ & $p$ \\
\midrule
Expression   & $+0.052$ & $<0.001$ & $+1.93$\,pp & $<0.001$ \\
Conditioning & $+0.012$ & $0.165$  & $+0.55$\,pp & $0.003$  \\
\midrule
Total        & $+0.063$ & $<0.001$ & $+2.47$\,pp & $<0.001$ \\
\bottomrule
\end{tabular}
\end{table}

Inspecting the user-agent scores explains the asymmetry. Conditioning raises the
mean score of \emph{all} candidates ($4.47{\to}5.37$) but lifts the ground-truth
item and the $49$ negatives almost equally, narrowing their margin
($1.07{\to}0.80$): personalized pitches make every candidate look better suited,
helping cold items in exposure allocation without sharpening the discrimination
the metric rewards. Replacing raw metadata with a promotion instead widens the
margin sharply ($0.31{\to}1.07$), as raw item memory mixes evaluative and
sometimes negative statements of inconsistent format.

\subsubsection{Can Items Game the Ranker by Exaggerating?}
We audit $6{,}000$ $(u,i)$ promotion pairs, oversampling exaggeration-suggestive
copy so that $74$ receive a \emph{strong} label ($243$ mild, $5{,}683$ none).
For each we take the user-internal standardised residual of the user-agent score
after removing a semantic-match baseline; a positive residual means the ranker
scored the item above its content similarity. There is no dose--response:
exaggeration is uncorrelated with the residual ($|\rho|{\le}0.015$, all
$p{>}0.23$), and mild exaggeration has no effect ($p{>}0.78$). Only strong
exaggeration shows one, and only under one of two baselines (TF-IDF $+0.41$,
$p{=}0.011$; embedding cosine $+0.16$, $p{=}0.31$). Crucially, the inflation does
not reach position: the rank gain of strongly exaggerated items is insignificant
($+0.76$, $p{=}0.75$). We thus report a bounded, not absent, gaming risk.

A fourth arm asks whether the gains are \emph{purchased} with embellishment.
\textbf{grounded} restricts generation to catalogue-verifiable attributes---swapping
the information source, not merely deleting adjectives from \textbf{full}.
Across CDs \& Vinyl and Movies \& TV ($|\mathcal{C}_u|{=}50$), an audit by
\texttt{claude-sonnet-4-6} of $200$ promotions per dataset raises the grounded-claim rate from
$78.0\%$/$79.5\%$ (\textbf{full}) to $96.0\%$/$94.5\%$ and cuts strong
exaggeration from $8.5\%$/$17.0\%$ to $0.5\%$/$2.0\%$. Yet the constraint costs
little: on zero-exposure items \textbf{grounded} retains $88.6\%$/$92.0\%$ of the
accuracy gain and $86.9\%$/$88.2\%$ of the exposure gain \textbf{full} obtains
over \textbf{none} (all gains over \textbf{none} significant, $p{<}0.001$), and
the \textbf{full}$-$\textbf{grounded} gap is insignificant on both datasets and
metrics (cold NDCG@10 $p{=}0.38$/$0.65$; exposure share $p{=}0.12$/$0.30$).
Embellishment is therefore not what produces the item-side gains.

\section{Conclusion and Limitations}

We proposed TriRec, a tri-party LLM-agent recommendation framework coordinating users, items, and the platform via item self-promotion and platform-led multi-objective re-ranking. TriRec outperforms baselines on tri-party utilities, and a controlled ablation attributes the gain to item-side expression---improving both ranking and zero-exposure items' exposure share---while user-conditioning adds exposure but no significant accuracy gain. Limitations: the offline protocol may not reflect long-term dynamics and self-promotion factuality is not yet human-validated, which future work targets via multi-round simulation, online A/B tests, retrieval-grounded constraints, and anti-gaming safeguards.

\section{Ethical Considerations}
\label{sec:ethics}

TriRec lets items generate persuasive, personalized self-promotion and lets a
platform agent redistribute exposure. Both capabilities raise concerns that we
make explicit here.

\paragraph{Hallucination and exaggeration.}
Because promotional copy is produced by an LLM, it may overstate or fabricate
item attributes. Our generation prompt instructs the agent to avoid exaggerated
marketing language and to ground every claim in the provided metadata. A
preliminary LLM-judged spot check indicates that this instruction alone is not
sufficient---invented format and edition attributes are the most common failure
mode. To check the automatic labels, one author manually re-annotated $100$
promotions (blind to the arm): the human labels agree with \texttt{claude-sonnet-4-6}
on $91\%$ of them (Cohen's $\kappa{=}0.59$, deflated by the high prevalence of
grounded copy), with near-perfect agreement on the \textbf{grounded} arm
($96\%$/$100\%$ on the two datasets) and the disagreements concentrated on
unconstrained \textbf{full} copy, where the automatic judge is if anything the
more lenient of the two. A larger multi-annotator audit remains necessary
before deployment, and deployments should retain a retrieval-grounded
verification filter and human review for high-stakes catalogs.

\paragraph{Adversarial self-promotion and gaming.}
If producers can influence the copy, they have an incentive to inflate appeal.
We do not evaluate this adversarial setting, so the robustness of the user agent
to deliberately exaggerated pitches remains an open question and a limitation of
this work. In practice, self-promotion should be treated as an untrusted signal,
combined with verifiable evidence, and subject to abuse detection.

\paragraph{Exposure redistribution and multi-stakeholder impact.}
Reallocating exposure toward long-tail producers benefits supply-side diversity
but can, if mis-calibrated, reduce short-term user relevance or over-promote
low-quality items. TriRec exposes this trade-off through an explicit,
inspectable objective rather than an opaque policy, and we report both accuracy
and fairness so the trade-off is transparent. A further caveat is that the fairness
reference is the historical training distribution, which itself encodes
popularity bias; operators adopting a different target distribution should
expect different exposure outcomes.

\paragraph{Reproducibility and cost.}
LLM scoring is stochastic (temperature $>0$) and incurs per-request API cost. We
disclose the sampling settings and backbone used for every reported result, and
report token consumption per request as a hardware-agnostic cost proxy.

\paragraph{Data.}
All experiments use publicly available benchmark datasets that contain no
personally identifiable information, and no human subjects were involved.

\bibliographystyle{ACM-Reference-Format}
\bibliography{references}

\end{document}